\documentclass[aps,prb,epsf,twocolumn,showpacs]{revtex4-2}

\usepackage{CJK,amsmath,amssymb,graphics,epsfig,epstopdf,color,verbatim,ulem,braket,tabularx,float}
\usepackage[colorlinks,linkcolor=blue,citecolor=blue,urlcolor=blue]{hyperref}

\begin{document}
\begin{CJK*}{GBK}{song}
\title{Spiral and Mixed Plaquette-Dimer Phases in the $S=1$ and $3/2$ Shastry-Sutherland Heisenberg Model}

\author{Han-Qing Wu$^{1}$}
\email{wuhanq3@mail.sysu.edu.cn}
\author{Muwei Wu$^{1,2}$}
\author{Shou-Shu Gong$^{3}$}

\affiliation{\mbox{$^{1}$Institute of Neutron Science and Technology,}
\mbox{Guangdong Provincial Key Laboratory of Magnetoelectric Physics and Devices,}
\mbox{School of Physics, Sun Yat-sen University, Guangzhou, 510275, China}
\mbox{$^{2}$Department of Physics, The Chinese University of Hong Kong, Shatin, New Territories,Hong Kong, China}
\mbox{$^{3}$School of Physical Sciences, Great Bay University, Dongguan 523000, China}
\mbox{Great Bay Institute for Advanced Study, Dongguan 523000, China}}

\begin{abstract}
  We investigate the ground-state phase diagram of the $S=1$ and $S=3/2$ Heisenberg model on the two-dimensional Shastry-Sutherland lattice (SSL) using density matrix renormalization group (DMRG) and cluster mean-field theory (CMFT). Between the dimer phase and N\'{e}el antiferromagnetic phases, we identify two intermediate phases: a mixed plaquette-dimer (MPD) phase and a spiral phase. These phases are characterized via bond energies and spin-spin correlation functions; phase boundaries are located from the ground-state energy derivative and entanglement entropy. The MPD phase exhibits strong intradimer correlations and weak tetramerization on the empty plaquettes, and its transitions to the dimer and spiral phases are first order. Combining our results with the known boundaries for $S=1/2$ and the classical limit $S\to\infty$, we construct a global $S$-$g$ phase diagram. This diagram reveals the progressive suppression of quantum effects with increasing $S$ and offers a theoretical framework for larger-$S$ SSL materials.
\end{abstract}

\date{\today}
\maketitle
\end{CJK*}

\textit{Introduction.} The Shastry-Sutherland lattice (SSL), formed by adding alternating diagonal couplings $J^\prime$ to a square lattice with nearest-neighbor coupling $J$, is a paradigmatic model of geometric frustration~\cite{SRIRAMSHASTRY19811069}. For the $S=1/2$ Heisenberg antiferromagnet with $g=J/J^\prime$, the phase diagram contains dimer, empty-plaquette (EP), and N\'{e}el antiferromagnetic (NAF) phases~\cite{PhysRevLett.84.4461,PhysRevB.64.134407,PhysRevB.66.014401,lou2012studyshastrysutherlandmodel,PhysRevB.87.115144,PhysRevB.92.094440,PhysRevX.9.041037,PhysRevB.100.140413,PhysRevB.105.L060409,PhysRevB.105.L041115,ChinPhysLett.39.077502,PhysRevLett.131.116702,PhysRevB.107.L220408,PhysRevLett.133.026502,PhysRevB.110.214410,chen2024spinexcitationsshastrysutherlandmodel,PhysRevB.111.134411,corboz2025quantumspinliquidphase,yuan2026spiralphasephasediagram}. Numerical studies also point to a possible quantum spin liquid (QSL) between the EP and NAF phases~\cite{PhysRevB.105.L060409,ChinPhysLett.39.077502,PhysRevB.105.L041115,PhysRevB.111.134411,corboz2025quantumspinliquidphase}. Even without a stable QSL, the EP-NAF transition remains intensely debated, with proposals ranging from a deconfined quantum critical point to a weakly first-order transition~\cite{doi:10.1126/science.1091806,PhysRevX.9.041037,PhysRevLett.133.026502, YCui2025,PhysRevB.87.115144,PhysRevB.107.L220408,chen2024spinexcitationsshastrysutherlandmodel, LChen2025}. In the classical limit $S\to\infty$, quantum order vanishes and a spiral phase emerges for $0<g<1$, transitioning into the NAF phase precisely at $g_c$=1~\cite{PhysRevB.72.104425,PhysRevB.79.144401}. Clarifying how this intricate phase diagram evolves from the extreme quantum $S=1/2$ regime to the classical limit is an open challenge.

\begin{figure}[htp]
  \centering
  \includegraphics[width=0.48\textwidth]{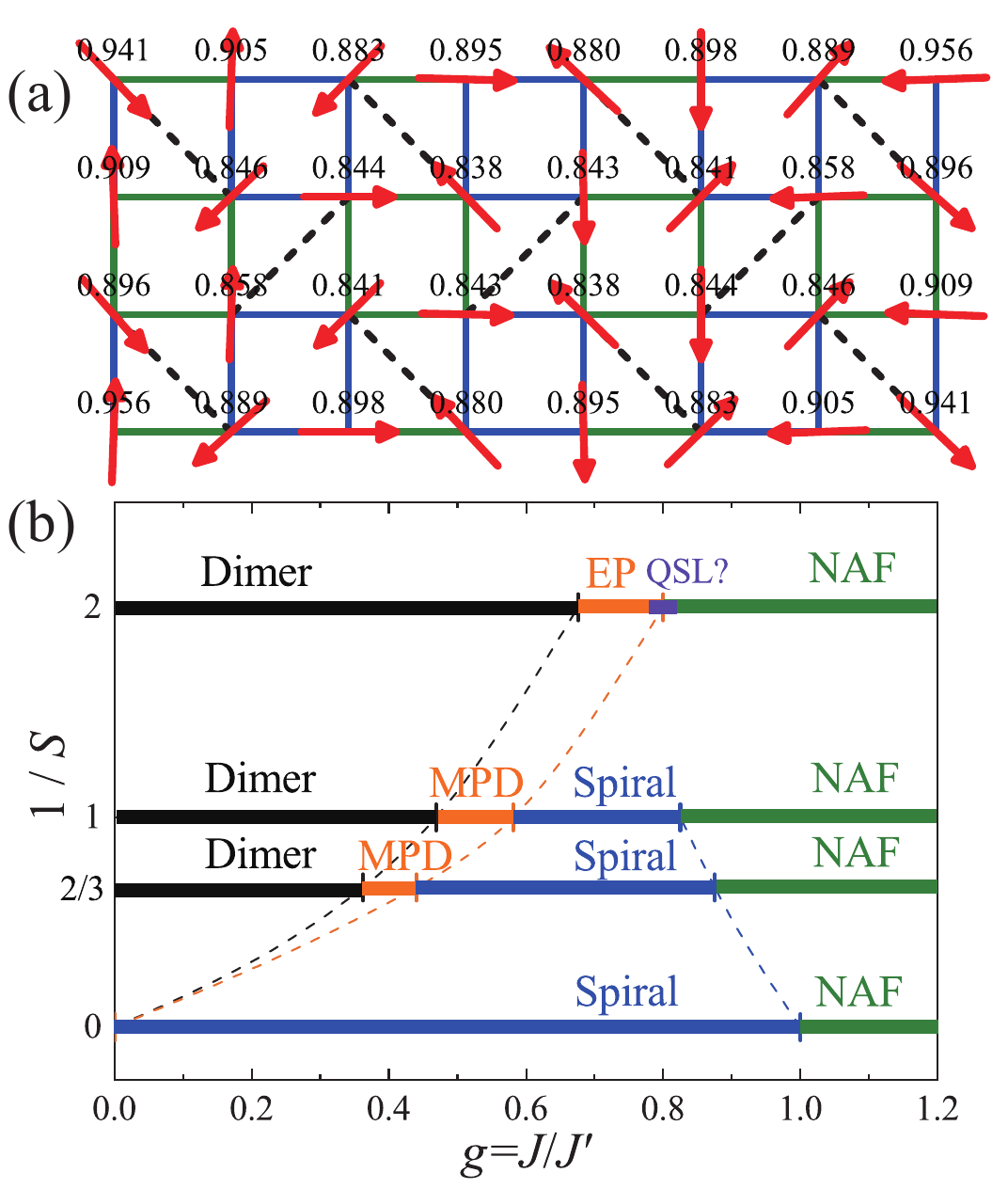}
  \caption{(a) Cylinder geometry with $L_y=4, L_x=8$. Solid (dashed) lines: $J$ ($J^\prime$) bonds. Empty plaquettes (without $J^\prime$) are colored to indicate two distinct bond-energy patterns (olive and blue). Red arrows show the normalized spin moments $\braket{\mathbf{S}_i}/S$ obtained by CMFT for $S=3/2$ at $g=1/\sqrt{2}$, with the magnitudes labeled. (b) Phase diagram as a function of $g=J/J^\prime$ for $S=1/2,1,3/2$, and the classical limit. Dashed lines are guides to the eye. The shaded region marks a possible quantum spin liquid (QSL) phase for $S=1/2$ (see text).}
  \label{fig:lattice}
\end{figure}

On the experimental side, SrCu$_2$(BO$_3$)$_2$ is the first material realization of the SSL, hosting $S=1/2$ moments on Cu$^{2+}$ ions~\cite{PhysRevLett.82.3168,PhysRevLett.82.3701,doi:10.1143/JPSJ.76.073710,doi:10.1073/pnas.1114464109,PhysRevLett.110.067210,PhysRevLett.113.067201,Haravifard2016,zayed2017,PhysRevLett.124.206602,Jimenez2021,Shi2022,Cui:2023jzi,cui2024plaquettesingletphasesshastrysutherlandcompound,Guo2025}. It enables the exploration of quantum criticality via external magnetic fields and pressure~\cite{chen2024spinexcitationsshastrysutherlandmodel,Cui:2023jzi,cui2024plaquettesingletphasesshastrysutherlandcompound,LilingSun2026}. More recently, new SSL compounds such as RE$_2$Be$_2$GeO$_7$(RE = Pr, Nd, Gd-Yb), ErB$_4$ and Eu$_2$MgSi$_2$O$_7$ have been synthesized~\cite{doi:10.1021/acs.inorgchem.0c03131,PhysRevB.110.014412,li2024spinonsnewshastrysutherlandlattice,PhysRevB.110.144445,ALiu2024,brassington2024,pula2025,m3wx-4v6k,gong2026}. Owing to strong spin-orbit coupling, the rare-earth ions can form effective $S=1/2$ doublets, carry larger spins $S$, or exhibit XYZ anisotropy~\cite{CLiu2025}. These developments call for a systematic understanding of larger-$S$ SSL physics. As a fundamental step, we investigate the isotropic Heisenberg model with larger $S$ on the SSL.


In this Letter, we determine the ground-state phase diagram of the isotropic Heisenberg model on the SSL for $S=1$ and $S=3/2$, using density matrix renormalization group (DMRG)~\cite{PhysRevLett.69.2863,RevModPhys.77.259} and cluster mean-field theory (CMFT) with DMRG as a solver~\cite{yuan2026spiralphasephasediagram}. We identify two intermediate phases between the dimer and NAF: a mixed plaquette-dimer (MPD) phase and a spiral phase, characterized via bond energies and spin-spin correlations. The spiral phase signals the evolution from the extreme quantum limit ($S=1/2$) toward the classical regime ($S\to\infty$): its stability region expands continuously with $S$, eventually covering the entire interval $0<g<1$ in the classical limit. The MPD phase exhibits strong intradimer correlations and weak tetramerization on the empty plaquette; as $S$ increases, its stability window shrinks and shifts to smaller $g$. These results bridge the gap between the well-studied $S=1/2$ and classical limits, providing a theoretical framework for larger-$S$ SSL materials.

\textit{Model and Method.} We consider the SSL depicted as a square lattice with alternating diagonal bonds [Fig.~\ref{fig:lattice}(a)]. The Heisenberg Hamiltonian is
\begin{equation*}
\begin{split}
H = J\sum\limits_{\left\langle {i,j} \right\rangle} {\mathbf{S}_{i} \cdot \mathbf{S}_{j}} + J'\sum\limits_{\left\langle\langle {i,j} \right\rangle\rangle^\prime} {\mathbf{S}_{i} \cdot \mathbf{S}_{j}},
\label{Eq:Hmlt}
\end{split}
\end{equation*}
where $J$ and $J^\prime$ denote the nearest-neighbor (inter-dimer) and diagonal (intra-dimer) couplings, depicted as solid and dashed lines in Fig.~\ref{fig:lattice}(a). We set $J^\prime=1$ as energy unit and define the coupling ratio $g=J/J^\prime$. Ground states are determined using DMRG, mostly on cylinders of width $L_y=4,6$ and length $L_x=24$, which can accommodate the incommensurate spiral correlations. Up to 8000 SU(2)-symmetric states are retained, giving a truncation error below $5\times10^{-6}$.
To characterize the spiral phase further, we employ cluster mean-field theory (CMFT) with DMRG as impurity solver, using the ITensor library~\cite{ITensor,yuan2026spiralphasephasediagram}. CMFT calculations use 
$L_x=8$ or $16$, $L_y=4$ clusters with open boundaries; intercluster interactions are treated at the mean-field level. The DMRG truncation error in these CMFT runs is kept below $10^{-6}$.


\begin{figure*}[htp]
  \centering
  \includegraphics[width=0.9\textwidth]{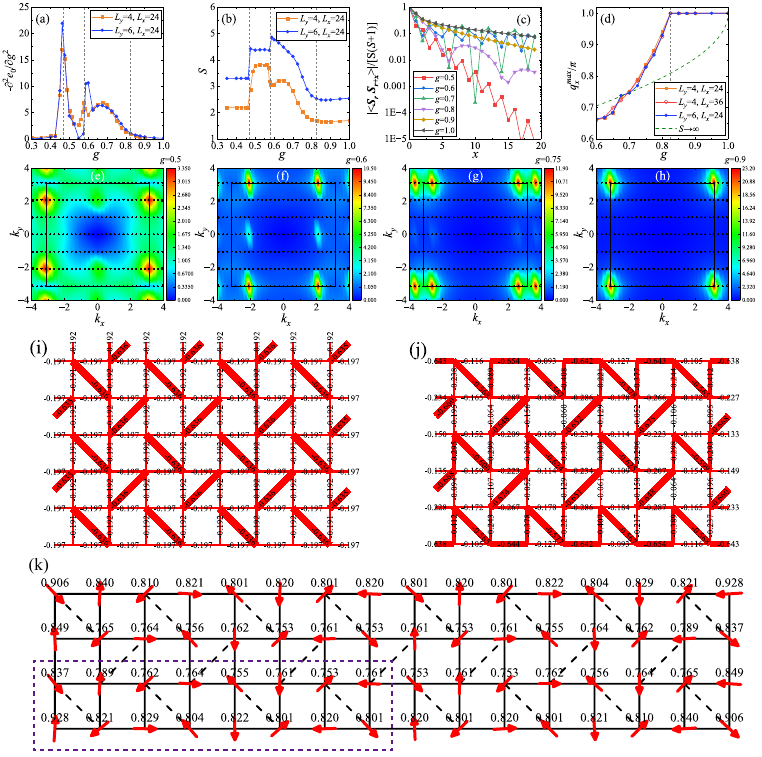}
  \caption{Numerical results for $S=1$. Vertical dashed lines in (a), (b), and (d) mark the phase transition points. (a) Second derivative of the ground-state energy per site as a function of the coupling ratio $g$. Orange and blue curves correspond to $L_y=4$ and $L_y=6$ cylinders, respectively. (b) Bipartite entanglement entropy $S_e$ versus $g$. (c) Real-space spin-spin correlation $| \langle\mathbf{S}_r\mathbf{S}_{r+x}\rangle |/[S(S+1)]$ along the $x$ direction for selected $g$ values. (d) Shift of the magnetic Bragg peak toward $M=(\pi,\pi)$ for $g\in[0.6,1.0]$. Orange and blue curves correspond to different cylinder widths, and the olive dashed line represents the classical limit $S\to\infty$. (e-h) Spin structure factors $S(\mathbf{q})$ at $g=0.5, 0.6, 0.75,$ and $0.9$, respectively. (i) Middle part of nearest-neighbor and next-nearest-neighbor bond energies at $g=0.5$ on the $L_y=6, L_x=24$ cylinder. (j) Same as (i), but with open boundary conditions in all directions, making the EP bond pattern evident. (k) Normalized spin moments $\braket{\mathbf{S}_i}/[S(S+1)]$ at $g$ = 1/$\sqrt{2}$ from CMFT on an $L_y$=4, $L_x$=16 cluster. The purple dashed box indicates the magnetic unit cell.}
  \label{fig:S1p0Main}
\end{figure*}

\textit{$S=1$ numerical results.} Figure~\ref{fig:S1p0Main} summarizes the DMRG results for $S=1$. The second derivative of the ground-state energy per site, $-\partial^2 e_0/\partial g^2$, shown in Fig.~\ref{fig:S1p0Main}(a), exhibits two sharp features at $g_{c1}\approx 0.47$ and $g_{c2}\approx 0.58$. These features, whose appear as discontinuities whose apparent divergence depends on the density of sampled $g$ points, signal first-order phase transitions. 
The bipartite entanglement entropy $S_e=-\text{tr}\rho_l \text{ln}\rho_l$ in Fig.~\ref{fig:S1p0Main}(b) independently confirms this picture: it displays clear jumps at both $g_{c1}$ and $g_{c2}$, making these two transitions more prominent. A broad peak in $-\partial^2 e_0/\partial g^2$ around $g\approx 0.675$ is likely a finite-size precursor of the transition at $g_{c2}$; its position shifts toward $g_{c2}$ as the system size grows, closely resembling the behavior reported for the $S=1$ orthogonal dimer chain model~\cite{PhysRevB.108.245104}. Thus, at least two quantum phase transitions occur within $g\in[0,1]$.

As in the $S=1/2$ case, the small-$g$ ground state is a dimer phase. At $g=0$, the exact ground state wave function is a product of dimer singlets $\prod_n \frac{1}{\sqrt{3}}(\ket{1,-1}_n-\ket{0,0}_n+\ket{-1,1}_n)$, where $n$ indexes diagonal dimers. For large $g$, the system enters the NAF phase. A natural expectation is that just above $g_{c1}$ an EP phase appears. To examine this intermediate regime, we focus on $g=0.5$ and show in Fig.~\ref{fig:S1p0Main}(e) the spin structure factor, together with the nearest- and next-nearest-neighbor bond energies on a $L_x=24, L_y=6$ cylinder (Fig.~\ref{fig:S1p0Main}(i)) and on a fully open cluster (Fig.~\ref{fig:S1p0Main}(j)). The spin structure factor reveals only short-range spin correlation for $g\in[g_{c1},g_{c2}]$. Unlike the $S=1/2$ EP phase, the bond energies display strong intradimer correlations accompanied by weak tetramerization on the empty plaquettes [Fig.~\ref{fig:S1p0Main}(j)]. These features identify a mixed plaquette-dimer (MPD) phase. Like the EP phase, the MPD phase has a two-fold degenerate ground state, corresponding to spin tetramers localized on one of the two empty plaquette sublattices [see olive and blue plaquettes in Fig.~\ref{fig:lattice}(a)]. In cylinder geometry, the superposition of the two degenerate states masks the weak tetramerization signal [Fig.~\ref{fig:S1p0Main}(i)]. Applying open boundary conditions lifts this degeneracy and makes the spin-tetramer character clearly visible in finite systems, as shown in Fig.~\ref{fig:S1p0Main}(j). The tendency toward tetramerization is also reflected in the real-space spin correlations along the $x$-direction at $g=0.5$ [Fig.~\ref{fig:S1p0Main}(c)], which exhibit a sawtooth-like exponential decay.

\begin{figure*}[htp]
  \centering
  \includegraphics[width=0.9\textwidth]{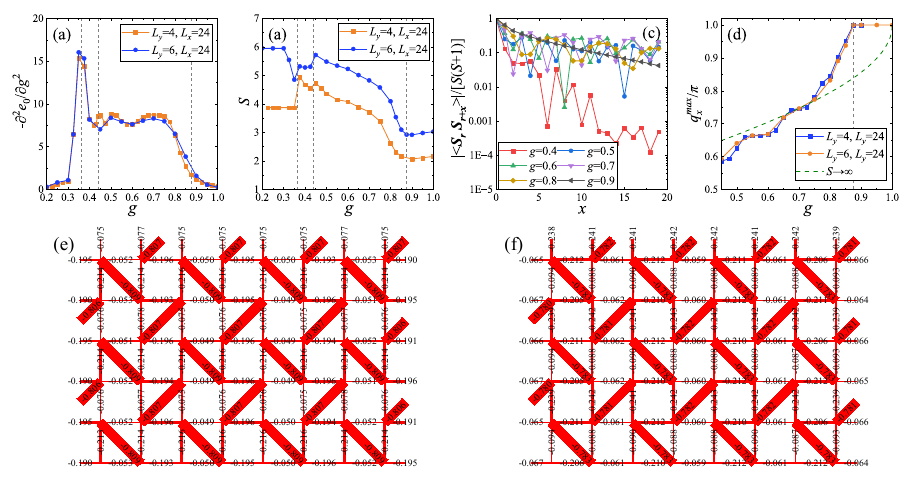}
  \caption{DMRG results for $S=3/2$. (a) Second derivative of the ground-state energy per site as a function of the coupling ratio $g$. Orange and blue curves correspond to $L_y=4$ and $L_y=6$ cylinders, respectively.
  (b) Bipartite entanglement entropy $S_e$ versus $g$. (c) Real-space spin-spin correlation $| \langle\mathbf{S}_r\mathbf{S}_{r+x}\rangle |/[S(S+1)]$ along the $x$ direction for selected values of $g$. (d) Shift of the magnetic Bragg peak toward $M=(\pi,\pi)$ for $g\in[0.45,1.0]$. Orange and blue curves correspond to different cylinder widths, and the olive dashed line represents the classical limit $S\to\infty$. (e), (f) Middle part of bond energies at $g=0.375$(e) and $g=0.4$(f) on an $L_y=6, L_x=24$ cylinder. Although both coupling ratios lie within the MPD phase, the two panels reflect the two degenerate tetramerization patterns.}
  \label{fig:S1p5Main}
\end{figure*}

For $g>g_{c2}$, the DMRG data show that the system does not immediately enter the NAF phase. Instead, it passes through a spiral phase characterized by a wave-vector $\mathbf{q}=(q_x,\pi)$ that varies with $g$, before eventually locking into the $\mathbf{q}=(\pi,\pi)$ NAF order. A similar phase was recently reported for the $S=1/2$ XXZ model on the SSL~\cite{yuan2026spiralphasephasediagram}. Figures~\ref{fig:S1p0Main}(e)-\ref{fig:S1p0Main}(h) display the static spin structure factor $S(\mathbf{q})=\frac{1}{N_s}\sum_{ij}e^{i\mathbf{q}(\mathbf{r}_j-\mathbf{r}_i)}\langle S_i S_j\rangle/[S(S+1)]$ at $g=0.5$, $0.6$, $0.75$, and $0.9$, respectively. A strong magnetic Bragg peak at $(q_x, \pi)$ is observed. For $g$ below a third critical value $g_{c3}$, $q_x$ is unlocked from $\pi$ and shifts continuously with $g$, a hallmark of incommensurate spiral order. We track this evolution by plotting $q_x / \pi$ versus $g$ in Fig.~\ref{fig:S1p0Main}(d). The wave vector reaches $\pi$ at $g_{c3}\approx 0.825$, which lies below the $g_{c3}=1$ in the $S\to\infty$ classical limit. Further support of $g_{c3}$ comes from the real-space spin correlations shown in Fig.~\ref{fig:S1p0Main}(c): in the spiral regime they decay as a power law with spatial oscillations, whereas for $g>g_{c3}$ they revert to a monotonic power-law decay characteristic of the NAF phase. Hence, the $S=1$ Heisenberg model on the SSL hosts two intermediate phases between dimer and NAF orders: an MPD phase and a spiral phase. Because of finite-size constraints in resolving incommensurate wave vectors, the spiral transition leaves no clear signature in the energy derivative [Fig.~\ref{fig:S1p0Main}(a)]. In contrast, the entropy $S_e$ in Fig.~\ref{fig:S1p0Main}(b) reveals it through a sharp drop followed by a plateau as $g$ increases.

To characterize the spiral order more directly, we apply CMFT with DMRG solver at the coupling ratio $g=1/\sqrt{2}$. This value is chosen because the $q_x-g$ curves for $S=1$ and the classical $S=\infty$ limit intersect there, yielding a spiral with a period-8 oscillation along $x$ for different spin magnitudes~\cite{PhysRevB.72.104425,PhysRevB.79.144401}. Figure~\ref{fig:S1p0Main}(k) presents the resulting spin texture: the directions of the magnetic moments display a clear 8$\times$2 periodicity, while the magnitudes show some deviations from perfect periodicity due to boundary effects.

We note that A. Koga \textit{et al.} previously studied the $S=1$ Heisenberg model on the SSL using exact diagonalization on a $4\times4$ torus~\cite{Koga2003}. Owing to the limited system size, the phase boundaries were not accurately determined, and the spiral phase, which requires larger systems, was not identified.

\textit{$S=3/2$ numerical results.}
Similar to the $S=1$ case, the $S=3/2$ Heisenberg model on the SSL exhibits two intermediate phases between the dimer and NAF states: an MPD phase and a spiral phase. The phase transitions are signaled by the second derivative of the energy and by the discontinuous jumps in the entanglement entropy at $g_{c1}\approx 0.36$ and $g_{c2}\approx 0.44$, as shown in Fig.~\ref{fig:S1p5Main}(a) and \ref{fig:S1p5Main}(b). Compared to $S=1$, an additional broad peak emerges in the second derivative of the energy near $g_{c3}$. The spiral order and the precise value 
$g_{c3} \approx 0.875$ are identified from the real-space and reciprocal-space spin correlations in Figs.~\ref{fig:S1p5Main}(c) and \ref{fig:S1p5Main}(d). For $g_{c2}<g<g_{c3}$, the magnetic Bragg peak $(q_x,\pi)$ shifts continuously to $(\pi,\pi)$ without any intervening phase transition. To examine the MPD phase, we present bond energies at $g=0.375$ and $0.4$ on a $L_x=24, L_y=6$ cylinder in Fig.~\ref{fig:S1p5Main}(e) and \ref{fig:S1p5Main}(f). Compared to $S=1$, the spin-tetramerized patterns on the two types of empty plaquettes are more pronounced, even on cylinder geometries, reflecting the two degenerate ground states. We further characterize the spiral order at $g=1/\sqrt{2}$ via CMFT on an $L_x=8$, $L_y=4$ cluster; the corresponding normalized spin moments $\braket{\mathbf{S}_i}/S$ are displayed in Fig.~\ref{fig:lattice}(a). These values are closer to unity than those obtained for $S=1$.

\textit{Phase diagram of different $S$.} Combining the phase boundaries determined here for $S=1$ and $S=3/2$ with the established results for $S=1/2$ and the classical limit $S\to\infty$, we construct the global $S-g$ phase diagram in Fig.~\ref{fig:lattice}(b). For $S=1/2$, we adopt the transition points $g_{c1}=0.676$ and $g_{c2}=0.8$ from Ref.~\cite{yuan2026spiralphasephasediagram}. The proposed quantum spin liquid (QSL) regime is indicated in purple (see, e.g., Refs.~\cite{PhysRevB.105.L060409,ChinPhysLett.39.077502,PhysRevB.105.L041115,PhysRevB.111.134411,corboz2025quantumspinliquidphase} and references therein). For $S\ge 1$, a spiral phase emerges between the MPD and NAF phases. As $S$ increases, the spiral region expands continuously, while the MPD phase shrinks and shifts toward smaller $g$ until it disappears; the dimer phase likewise diminishes and vanishes in the classical limit. In the $S\to\infty$ limit, the phase diagram reduces to the classical case, with only the spiral and NAF phases. These results show that even for larger spins, quantum fluctuations combined with geometric frustration can stabilize exotic phases such as the MPD phase, which have no classical counterpart.

\textit{Summary and discussion.} Using DMRG and CMFT, we have investigated the ground-state phase diagram of the Heisenberg model on the Shastry-Sutherland lattice for $S=1$ and $3/2$. We identify two intermediate phases between the dimer and NAF phases: a mixed plaquette--dimer (MPD) phase, characterized by strong intradimer correlations and weak tetramerization on the empty plaquettes, and a spiral phase with an incommensurate wave vector $(q_x,\pi)$ that shifts continuously to $(\pi,\pi)$ with increasing $g$. Combining our results with the known phase boundaries for $S=1/2$ and the classical limit ($S\to\infty$), we construct a unified $S$-$g$ phase diagram spanning from the quantum to classical regimes. 

Our results demonstrate that quantum fluctuations and geometric frustration can stabilize nonmagnetic phases, such as the MPD state, even for $S=3/2$. The spiral phase for $S\ge 1$ bridges the quantum-disordered ground state at $S=1/2$ and the classical spiral order. Given the recent synthesis of rare-earth-based Shastry-Sutherland materials with effective larger spins, this phase diagram provides a theoretical framework for interpreting experiments and predicting magnetic ground states in these compounds.

\begin{acknowledgments}
H.Q.W. and M.W. are supported by the National Natural Science Foundation of China(Grant No. 12474248),Guangdong Basic and Applied Basic Research Foundation (Grant No. 2023B1515120013), Guangdong Provincial Key Laboratory of Magnetoelectric Physics and Devices (No. 2022B1212010008) and Guangdong Fundamental Research Center for Magnetoelectric Physics (Grant No.2024B0303390001). 
S. S. G. was supported by the National Natural Science Foundation of China (No. 12274014 and No. 12534009), the Guangdong Provincial Quantum Science Strategic Initiative (Grant No. GDZX2501006), the Special Project in Key Areas for Universities in Guangdong Province (No. 2023ZDZX3054), and the Dongguan Key Laboratory of Artificial Intelligence Design for Advanced Materials.
\end{acknowledgments}

\bibliography{SSL_HS}

\end{document}